\documentclass[a4paper,12pt]{article}
\usepackage[utf8]{inputenc}

\usepackage{graphicx}
\usepackage{amsmath}
\usepackage{amsfonts}
\usepackage{amssymb}
\usepackage{amsthm}

\usepackage{hyperref}
\newtheorem{definition}{Definition}

\usepackage{amsmath}

\title{Activity of any neuron with delayed feedback stimulated with Poisson stream is non-Markov}\author{Alexander K.Vidybida
\thanks{Bogolyubov Institute for Theoretical Physics,
Metrologichna str., 14-B, Kyiv 03680, Ukraine,
vidybida@bitp.kiev.ua,
http://www.bitp.kiev.ua/pers/vidybida}
}
\date{}

\begin{document}
\renewcommand{\abstractname}{Abstract}
\renewcommand{\refname}{References}
\renewcommand{\figurename}{Fig}
\maketitle
\thispagestyle{empty} 

\begin{abstract}
For a class of excitatory spiking neuron models with delayed feedback
 fed with a Poisson stochastic process, it is proven that
the stream of output interspike intervals cannot be presented as a Markov process of any order.


\end{abstract}





\section{Introduction}

Statistics of neuronal activity is often described as a renewal point process, or even a Poisson process,
see \cite{Averbeck2009} and references therein. On the other hand, in some sets of experimental data 
correlations are observed between consecutive interspike intervals (ISI), \cite{Lowen1992,Ratnam2000,Nawrot2007,Maimon2009}, which does not conform with the renewal hypothesis.
What could be the reason of such correlations? In principle, any sort of memory in the neuronal firing 
mechanism could bring about memory into the sequence of ISIs, thus disrupting a possibility for it to be renewal. Memory in the firing mechanism can appear due to partial reset of the membrane potential
after firing, \cite{Rospars1993a,Lnsk1999}, or due to threshold fatigue \cite{Chacron2003},
or for other reasons, see \cite{Avila-Akerberg2011} for a review. 

Biologically, non-renewal statistics of neuronal activity can improve discrimination of weak signals
\cite{Ratnam2000,Avila-Akerberg2011} and therefore is essential feature of functioning of a nervous system.
In this context, it was checked in \cite{Ratnam2000} if it is possible to represent activity of
 electrosensory neuron as a Markov chain of some finite order. Conclusion made in \cite{Ratnam2000} is
 that the corresponding order, if any, cannot be lower than 7.

Normally, any neuron is embedded into a network. Inter-neuronal communication in the network is delayed
due to finite speed of nervous impulses. In a reverberating network, this brings about one more reason
for non-renewal firing statistics --- the delayed feedback. We study here the simplest possible case of
a network --- a single neuron with delayed feedback. In the previous paper \cite{Vidybida2012}, it was proven
for a concrete neuronal model --- the binding neuron with threshold 2 --- stimulated with Poisson stream
of input impulses, that statistics of its ISIs is essentially non-Markov. In this paper, we refine 
and extend methods
of \cite{Vidybida2012} making those applicable to any neuron, which satisfies a number of very simple and
natural conditions (see Cond0-Cond4 in n. \ref{neuron}). Under those conditions, we prove rigorously that
ISI statistics of a neuron with delayed feedback cannot be represented as a Markov chain of any finite
order.

\section{Definitions and assumptions}

\subsection{Neuron without feedback}\label{neuron}

We do not specify any concrete neuronal model, only expect that a neuron
satisfies the following conditions:
\begin{itemize}
\item Cond0: Neuron is deterministic: Identical stimuli elicit identical spike trains from the same neuron.
\item Cond1: Neuron is stimulated with input Poisson stream of excitatory impulses.
The input stream has intensity $\lambda$.
\item Cond2: Neuron may fire a spike only at a moment when it receives an input impulse.
\item Cond3: Just after firing, neuron appears in its standard state, which is always the same.
\item Cond4: The output interspike interval (ISI) distribution is characterized with a
probability density function (pdf) $p^0(t)$, which is positive: $t>0\Rightarrow p^0(t)>0$, and
bounded: $\sup\limits_{t>0} p^0(t)< \infty$.
\end{itemize}

The Cond0, above, is imposed in accordance with experimental observations, 
see e.g. \cite{Bryant1976,Mainen1995}.
As regards the Cond1,  Poisson stream is a standard stimulation when neuronal random activity is studied.
 The Cond2, above, is satisfied for most threshold-type neuronal models, starting from 
standard leaky integrate and fire (LIF) neuron \cite{Stein1967} and its modifications, see \cite{Burkitt}.
In order the Cond2 to be valid, it is enough that the following three 
conditions are satisfied: (i) neuronal excitation%
\footnote{We use here term ``excitation'' instead of ``depolarization voltage'' because we do not specify
any triggering mechanism. Our consideration as regards feedback shaping of firing statistics could be 
valid also for essentially artificial neurons, where excitation not necessarily has a voltaic nature.}
 gets abrupt increase at the moment of receiving input impulse%
\footnote{If considering an input impulse as a current impulse, then it has a $\delta$-function form.}, 
(ii) after that moment, the degree of excitation does not increase 
(it decreases for most neuronal models) until the next input impulse. 
(iii) the neuron fires when its degree of excitation exceeds a threshold level. The threshold can be either
static, as in the basic LIF model, or dynamic \cite{Segundo1968}. These conditions seem to be
standard for many threshold neuronal models used,  see \cite{Chacron2003,Jolivet2004,Jolivet2006} and citations therein. Cond3 means that any kind of memory about previous input/output activity, 
which can be present in a neuron, is cleared
 after each triggering. Due to Cond3, output stream of neuron without feedback will be a renewal
 stochastic process.
Cond4 seems to be natural for any neuronal model stimulated with Poisson stream. At least, 
all the five conditions are satisfied for the binding neuron model and for the basic LIF model, see \cite{Vidybida2007,Vidybida2014b}, where $p^0(t)$ is calculated exactly for each model, respectively.

\subsection{Feedback line action}\label{line}

We expect that each output impulse fired by neuron is fed back to the neuron's input
through a feedback line. The feedback line has the following properties:
\begin{itemize}
\item Prop1: The time delay in the line is $\Delta>0$.
\item Prop2: The line is able to convey no more than one impulse.
\item Prop3: The impulse conveyed to the neuronal input is identical to that from the input 
Poisson stream.
\end{itemize}

It is known that a neuron can form synapses (autapses) on its own body, or dendritic tree, e.g.
\cite{Nicoll1982,Bekkers1998}.
This substantiates consideration of a single neuron with feedback not only as the simplest reverberating
"network" possible, bat also as an independent biologically relevant case.
 The delay $\Delta$ comprises the time required by the output spike to 
pass the distance from axonal hillock, where it is generated, to the autapse and the synaptic delay.
The Prop2 is somehow related to the refractoriness even if we do not introduce here the refractoriness 
to its full extent. 
The Prop3 means that we consider here an excitatory neuron.

The important for us consequence of Prop2 is that at any moment of time the feedback
line is either empty, or conveys a single impulse. If it does convey an impulse, then its state
can be described with a stochastic variable $s$, which we call further ``time to live''. 
The variable $s$ denotes the exact time required by the impulse to reach the output end of the line,
which is the neuron's input, and to leave the line. It is clear that $0<s\le\Delta$. In what follows,
we use the time to live $s$ only at moments when an ISI starts (just after triggering).

Now it is worth to notice that each triggering starts a new ISI. 
And at the beginning of any ISI the line is never empty, but
 holds an impulse. This happens for the following reasons:
\begin{itemize}
\item[a)] If neuron is triggered by an impulse from the Poisson input stream, and the line was
empty just before that moment, then the emitted impulse enters the line. 
At that moment the line is characterized with $s=\Delta$.
\item[b)] If neuron is triggered by an impulse from the Poisson input stream, and the line already
conveys an impulse at that moment with time to live $s$, then that same impulse with that same 
time to live is retained at the beginning of the ISI that starts after that triggering, and the line is 
characterized with that same $s$.
\item[c)] If neuron is triggered by an impulse from the line, then the line is empty at the firing moment
and the emitted impulse enters the line. After that moment the line is characterized
with $s=\Delta$.
\end{itemize}

\subsection{Proof outline}
We expect that defined in nn. \ref{neuron}, \ref{line} system of neuron with delayed
feedback line fed with Poisson stream is in its stationary regime. This can be achieved
if the system functions long enough that its initial state is forgotten.

In the stationary regime, let $p(t_n,\dots,t_1)$ denotes the 
joint probability density function of neuron with delayed feedback.
The probability to get, in the output, starting from the beginning, $n$
consecutive ISIs $t'_1,\dots,t'_n$ such that $t'_i\in[t_i;t_i+dt_i[$, $i=1,\dots,n$
with infinitesimal $dt_i$ is given by $p(t_n,\dots,t_1)dt_1\dots dt_n$.

Let $p(t_{n+1}\mid t_n,\dots,t_0)dt_{n+1}$ denotes the conditional probability 
to get the duration of $(n+2)$-th ISI in $[t_{n+1};t_{n+1}+dt_{n+1}[$ provided 
that previous $n+1$ ISIs had duration $t_n,\dots,t_0$, respectively.

Now we reformulate in terms of probability density functions
 the definition from \cite[Ch.2 \S 6]{Doob1953}:
\begin{definition}
The sequence of random variables $\{t_{j}\}$, taking values in $\Omega$, is called the Markov chain of the order $n\ge0$, if
\begin{displaymath}
        \forall_{m > n} \forall_{t_0\in\Omega}\ldots \forall_{t_m\in\Omega}\
        p(t_{m}\mid t_{m-1},\ldots,t_{0})
        = p(t_{m}\mid t_{m-1},\ldots,t_{m-n}),
\end{displaymath}
and this equation does not hold for any $n'<n$.
\end{definition}

In particular, taking $m=n+1$, we have the necessary condition
\begin{multline}
\label{def}
        p(t_{n+1}\mid t_{n},\ldots,t_{1},t_{0})
        = p(t_{n+1}\mid t_{n},\ldots,t_{1}),~~
        t_i\in\Omega,~  i=0,\ldots,n+1,
\end{multline}
required for the stochastic process $\{t_{j}\}$ to be $n$-order Markov chain.
In the case of ISIs one reads $\Omega=\mathbb{R^+}$.

We intend to prove that the relation (\ref{def}) does not hold for any $n$. For this purpose
we calculate exact expression for $p(t_{n+1}\mid t_n,\dots,t_0)$ as
\begin{equation}\label{defcond}
p(t_{n+1}\mid t_n,\dots,t_0)=
\frac{p(t_{n+1}, t_n,\dots,t_0)}{p(t_n,\dots,t_0)}
\end{equation}
from which it will be clearly seen that the $t_0$-dependence in $p(t_{n+1}\mid t_n,\dots,t_0)$
cannot be eliminated whatever large the $n$ is. 

As it is seen from (\ref{defcond}), we need initially to calculate exact expressions 
for $p(t_n,\dots,t_0)$ with arbitrary $n$. In \cite{Vidybida2012}, for the binding
neuron model with threshold 2 
this is done by introducing an auxiliary stochastic process with events $(t_i,s_i)$,
where $s_i$ is the time to live at the beginning of ISI $t_i$. It was proven that
the sequence of events 
$(t_i,s_i)$, $i=0,1,\dots$, is Markov chain, which helps to calculate the joint probability density
$p((t_n,s_n),\dots,(t_0,s_0))$ and then $p(t_n,\dots,t_0)$ as marginal probability
by integrating it over $]0;\Delta]$ with respect to each $s_i$. To simplify this
approach, it is worth to notice that in the sequence of consecutive random events
$(t_n,s_n),\dots,(t_0,s_0)$ only the values of variables
$t_n,\dots,t_1,t_0,s_0$ are fairly random. Indeed,
with $t_0,s_0$ given, one can figure out exact value for the $s_1$: if $t_0<s_0$ then
$s_1=s_0-t_0$, and $s_1=\Delta$ otherwise. Now, with $t_1,s_1$ known, the same way it is
possible to find the exact value of $s_2$ and so on. This allows one to reconstruct unambiguously
all the values $s_1,\dots,s_n$ from the given sequence of values of $t_n,\dots,t_1,t_0,s_0$.
Having this in mind, we introduce the conditional joint probability density
$p(t_{n+1},\dots,t_0\mid s)$, which we use to calculate required joint pdfs as follows
\begin{equation}\label{pdfs}
p(t_{n+1},\dots,t_0)=\int\limits_0^\Delta p(t_{n+1},\dots,t_0\mid s)f(s)\,ds,
\end{equation}
where $s$ (dented previously as $s_0$) is the time to live at the beginning of ISI $t_0$, 
$f(s)$ is the stationary pdf which describes distribution of times to live
at the beginning of any ISI
in the stationary regime. In what follows we analyze the structure of functions
$f(s)$ and $p(t_{n+1},\dots,t_0\mid s)$. It appears that $f(s)$ has a singular
component $a\delta(s-\Delta)$ with $a>0$, and $p(t_{n+1},\dots,t_0\mid s)$
has a $\delta$-function-type singularities at definite hyper-planes in the $(n+2)$-dimensional
space of its variables $(t_{n+1},\dots,t_0)$. 
 After integration in (\ref{pdfs}), some of those $\delta$-functions
will survive, and one of those survived has its argument depending on $t_0$. 
The latter statement depends on exact value of ISIs in the sequence $t_{n+1},\dots,t_0$.
Here, we limit our consideration to the domain in the $(n+2)$-dimensional
space of variables $(t_{n+1},\dots,t_0)$, which is defined as follows
\begin{equation}\label{Domain}
\sum\limits_{i=0}^n t_i<\Delta.
\end{equation}
Notice that $t_{n+1}$ is not involved in (\ref{Domain}).

The $t_0$-dependent $\delta$-function
will as well survive in the $p(t_{n+1}\mid t_n,\dots,t_0)$ for any $n$, which will complete
the proof that the condition (\ref{def}) cannot be satisfied for any $n$.

A question remains of whether the domain (\ref{Domain}) has a strictly positive probability.
This indeed takes place due to positiveness of pdfs $p(t_{n+1},\dots,t_0)$ for any positive values of 
$(t_{n+1},\dots,t_0)$. The latter follows from the exact expressions for $p(t_{n+1},\dots,t_0)$
given in n. \ref{Form}, Eq. (\ref{pn}).

\section{The proof}

\subsection{Structure of functions $p(t_{n+1},\dots,t_0 \mid s)$}
Expect that the inequality (\ref{Domain}) holds.
In order to perform integration in (\ref{pdfs}), we split the integration domain 
into the following $n+2$ disjoint sub-domains:
$$
D_k=\left]\sum\limits_{i=0}^{k-1}t_i\,;~\sum\limits_{i=0}^{k}t_i\right],~
k=0,\dots,n,~~
D_{n+1}=\left]\sum\limits_{i=0}^{n}t_i\,;~\Delta\right]\,.
$$
It is clear that
$$
\bigcup\limits_{k=0}^{n+1}D_k = ]0;\Delta].
$$
\begin{figure}[h]
\begin{center}
\includegraphics[width=0.66\textwidth,angle=0]{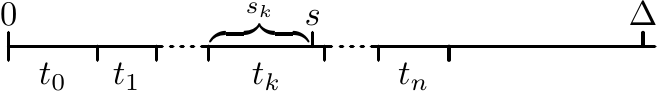}
\end{center}
\caption{\label{relk}Mutual disposition in time of $s$ and $t_0,\dots,t_n$ if $s\in D_k$.}
\end{figure}
The conditional pdf $p(t_{n+1},\dots,t_0 \mid s)$ has different structure at different domains.
If $s\in D_k$, then a relation between $s$ and $t_i$ is as shown in Fig. \ref{relk}.
As it could be suggested by Fig. \ref{relk}, the first $k-1$ ISIs are produced with the delay
line not involved. The $k$-th ISI is generated with the line involved. The corresponding
time to live is $s_k=s-\sum_{i=0}^{k-1} t_i\le t_k$, the next time to live is $s_{k+1}=\Delta$.
Therefore, the structure of $p(t_{n+1},\dots,t_0 \mid s)$  at $D_k$ is as follows
\begin{equation}\label{struk}
p(t_{n+1},\dots,t_0 \mid s)=p(t_{n+1},\dots,t_{k+1} \mid \Delta)\,
p\left(t_k \mid s-\sum\limits_{i=0}^{k-1} t_i\right)
\prod\limits_{i=0}^{k-1}p^0(t_i),
\end{equation}
\begin{figure}[h]
\begin{center}
\includegraphics[width=0.66\textwidth,angle=0]{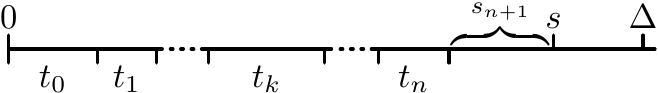}
\end{center}
\caption{\label{reln}Mutual disposition in time of $s$ and $t_0,\dots,t_n$ if $s\in D_{n+1}$.}
\end{figure}
where $k=0,1,\dots,n$. And if $s\in D_{n+1}$, then relation between $s$ and $t_i$
is as shown in Fig. \ref{reln}. This suggests the following structure for 
$p(t_{n+1},\dots,t_0 \mid s)$
\begin{equation}\label{strun+1}
p(t_{n+1},\dots,t_0 \mid s)=
p\left(t_{n+1} \mid s-\sum\limits_{i=0}^{n} t_i\right)
\prod\limits_{i=0}^{n}p^0(t_i),~ s\in D_{n+1}.
\end{equation}
Here $p(t\mid s)$ denotes the conditional pdf to get ISI of duration $t$ if
at its beginning, time to live of impulse in the feedback line is $s$. 

By utilizing the same reasoning with (\ref{Domain}) taken into account, 
one can represent the first factor in (\ref{struk})
as follows
\begin{equation}\label{strukf}
p(t_{n+1},\dots,t_{k+1} \mid \Delta)=
p\left(t_{n+1} \mid \Delta-\sum\limits_{i=k+1}^{n} t_i\right)
\prod\limits_{i=k+1}^{n}p^0(t_i).
\end{equation}

Representation of $p(t_{n+1},\dots,t_0 \mid s)$ by means of $p^0(t)$ and $p(t\mid s)$,
similar to that displayed in (\ref{struk}), (\ref{strun+1}), (\ref{strukf}), 
can be as well constructed if (\ref{Domain}) does not hold.
For our purpose it is enough to have (\ref{struk}), (\ref{strun+1}) and (\ref{strukf}).

\subsection{Structure of function $p(t \mid s)$}

Expect that at the beginning of an ISI, there is an impulse in the 
feedback line with time to live $s$. Then the probability that this ISI
will have its duration $t<s$ does not depend on the feedback
line presence. Therefore,
\begin{equation}\label{pinit}\nonumber
t<s\, \Rightarrow\, p(t\mid s) = p^0(t).
\end{equation}
The probability to get exactly $t=s$ is not zero, because in this case
the impulse, which triggers the neuron and finishes the ISI
under consideration comes from the delay line. In order
this to happen, it is necessary and sufficient that the following two
events take place: (i) the neuron does not fire at the interval $]0;s[$;
(ii) at the moment $s$, the neuron, due to previous stimulation from the Poisson stream, achieves
such a state that adding one more input impulse will trigger it.
The probability of (i) and (ii) is $\frac{p^0(s)}{\lambda}$, which can be easily concluded  
from the definition of $p^0(t)$. Thus,
\begin{equation}\label{pd}\nonumber
t\in\,\,]s-\epsilon;s+\epsilon[\,\, \Rightarrow\, p(t\mid s) = \frac{p^0(t)}{\lambda}\delta(s-t)
\end{equation}
with infinitesimal $\epsilon>0$. If the neuron still not triggered at moment $s$,
then it is triggered by an input impulse from the Poisson stream at $t>s$. The probability to
get such an impulse in $[t;t+dt[$ is $\lambda dt$. Therefore, one can expect
that for $t>s$, $p(t\mid s)\le\lambda$.

Based on the above reasoning we represent $p(t\mid s)$ in the following form
\begin{equation}\label{pstruc}
p(t\mid s) = p^b(t\mid s) + \frac{p^0(t)}{\lambda}\delta(s-t),
\end{equation}
where $p^b(t\mid s)$ is a bounded function\footnote{Compare this with \cite[Eq. (7)]{Vidybida2008a}, where $p(t\mid s)$ is calculated exactly for the binding neuron model.}.

\subsection{Structure of probability density function $f(s)$}

In the stationary regime, the pdf $f(s)$ must satisfy the following equation
\begin{equation}\label{trans}
f(s) = \int\limits_0^\Delta \mathbf{P}(s\mid s') f(s') ds',
\end{equation}
where the transition function $\mathbf{P}(s\mid s')$ gives the probability density
to find at the beginning of an ISI an impulse in the line with time to live $s$
provided at the beginning of the previous ISI, there was an impulse with time to live $s'$.

To determine exact expression for $\mathbf{P}(s\mid s')$ we take into account that
after single firing, time to live can either decrease, or become equal $\Delta$.
Therefore,
\begin{equation}\label{trans1}
s'\le s <\Delta\,\,\Rightarrow \, \mathbf{P}(s\mid s')=0.
\end{equation}
If $s<s'$, then the firing, which causes transition from $s'$ to $s$, happens
without the line involved. Therefore,
\begin{equation}\label{trans2}
0 < s < s'\,\,\Rightarrow \, \mathbf{P}(s\mid s')ds = p^0(s'-s)ds.
\end{equation}
Finally, it is possible that starting from $s'$ one obtains $s=\Delta$ after
the next firing. In order this to happen, it is necessary and sufficient that
no firing happens during $s'$ units of time. And this happens with probability
$$
\mathbf{P}^0(s') = 1 - \int\limits_0^{s'} p^0(t)dt.
$$
Having this in mind, one could conclude that in the plane $(s,s')$, at the straight line $s=\Delta$,
$s'$ --- any, 
the $\mathbf{P}(s\mid s')$ has singularity of the following form:
\begin{equation}\label{trans3}
\mathbf{P}^0(s')\delta(s-\Delta).
\end{equation}

Now, with (\ref{trans1})-(\ref{trans3}) taken into account, Eq. (\ref{trans})
can be rewritten as follows
\begin{equation}\label{feq}\nonumber
f(s)=\int\limits_s^\Delta p^0(s'-s)f(s')ds' +
\delta(s-\Delta)\int\limits_0^\Delta \mathbf{P}^0(s') f(s')ds'.
\end{equation}
It is clear from this equation that $f(s)$ has the following form\footnote{Compare this with \cite[Eqs. (14)-(16)]{Vidybida2008a}, where $f(s)$ is calculated exactly for the binding neuron model.}
\begin{equation}\label{ff}
f(s)=g(s) + a\delta(s-\Delta),
\end{equation}
where $a>0$ and $g(s)$ is bounded and vanishes out of interval $]0;\Delta]$.

\subsection{Form of $p(t_{n+1},\dots,t_0)$ and $p(t_{n},\dots,t_0)$ after integration in (\ref{pdfs})} 

Let $D=\bigcup\limits_{k=0}^n D_k$. 
At $D$, representations (\ref{struk}) and (\ref{strukf}) are valid. Also at $D$,
$f(s)$ reduces to $g(s)$.
Therefore,
\begin{multline}\label{intD}
\int\limits_D p(t_{n+1},\dots,t_0\mid s) f(s) =
\\=
\sum\limits_{k=0}^np\left(t_{n+1}\mid \Delta -\sum\limits_{i=k+1}^n t_i\right)
\prod\limits_{\vbox{\footnotesize\hbox{$i=0$}\hbox{$i\ne k$}}}^np^0(t_i)
\int\limits_{D_k}p\left(t_k\mid s-\sum\limits_{i=0}^{k-1} t_i\right) g(s)ds.
\end{multline}
Taking into account (\ref{pstruc}) it can be concluded that expression (\ref{intD}),
after performing integration, does not have any term with $\delta$-function 
depending on $t_0$.

Consider now the remaining part of integral in (\ref{pdfs}). With (\ref{strun+1})
taken into account one has:
$$
\int\limits_{D_{n+1}} p(t_{n+1},\dots,t_0\mid s) f(s) 
=
\prod\limits_{i=0}^n p^0(t_i)
\int\limits_{D_{n+1}}p\left(t_{n+1}\mid s-\sum\limits_{i=0}^{n}t_i\right) f(s)ds.
$$
After substituting here expressions (\ref{pstruc}), (\ref{ff}) one obtains four terms:
\begin{multline}\label{intDn+1}
\int\limits_{D_{n+1}} p(t_{n+1},\dots,t_0\mid s) f(s) =
\\=
\prod\limits_{i=0}^n p^0(t_i)
\int\limits_{D_{n+1}}p\left(t_{n+1}\mid s-\sum\limits_{i=0}^{n}t_i\right) f(s)ds
\\=
\prod\limits_{i=0}^n p^0(t_i)
\int\limits_{D_{n+1}}p^b\left(t_{n+1}\mid s-\sum\limits_{i=0}^{n}t_i\right) g(s)ds
\\+
a \prod\limits_{i=0}^n p^0(t_i)
 p^b\left(t_{n+1}\mid \Delta-\sum\limits_{i=0}^{n}t_i\right)
\\+
\frac{1}{\lambda}\prod\limits_{i=0}^{n+1} p^0(t_i) g\left(\sum\limits_{i=0}^{n+1}t_i\right)
+
\frac{a}{\lambda}\prod\limits_{i=0}^{n+1} p^0(t_i) 
\delta\left(\Delta-\sum\limits_{i=0}^{n+1}t_i\right).
\end{multline}
After performing integration, only the fourth term here includes a $\delta$-function.
And argument of this $\delta$-function does depend on $t_0$.

After taking (\ref{intD}) and (\ref{intDn+1}) together we conclude that the required joint
probability density has the following form
\begin{equation}\label{pn+1}
p(t_{n+1},\dots,t_0)=p^w(t_{n+1},\dots,t_0)+
\frac{a}{\lambda}\prod\limits_{i=0}^{n+1} p^0(t_i) 
\delta\left(\Delta-\sum\limits_{i=0}^{n+1}t_i\right),
\end{equation}
where function $p^w(t_{n+1},\dots,t_0)$ does not have singularities depending on $t_0$.

\subsubsection{Form of $p(t_{n},\dots,t_0)$ after integration}\label{Form}
If (\ref{Domain}) is satisfied, then we have similarly to (\ref{struk}), (\ref{strun+1})
\begin{multline}\nonumber
p(t_{n},\dots,t_0 \mid s)=p(t_{n},\dots,t_{k+1} \mid \Delta)\,
p\left(t_k \mid s-\sum\limits_{i=0}^{k-1} t_i\right)
\prod\limits_{i=0}^{k-1}p^0(t_i),
\\
 s\in D_k,\quad k=0,\dots,n-1,
\end{multline}
$$
p(t_{n},\dots,t_0 \mid s)=
p\left(t_n \mid s-\sum\limits_{i=0}^{n-1} t_i\right)
\prod\limits_{i=0}^{n-1}p^0(t_i),\quad s\in D_n.
$$
Again due to (\ref{Domain}), and in analogy with (\ref{strukf}) we have instead 
of the last two equations the following one:
\begin{multline}\label{strn}
p(t_{n},\dots,t_0 \mid s)=
p\left(t_k \mid s-\sum\limits_{i=0}^{k-1} t_i\right)
\prod\limits_{\vbox{\footnotesize\hbox{$i=0$}\hbox{$i\ne k$}}}^{n}p^0(t_i),
\\
 s\in D_k,\quad k=0,\dots,n.
 \end{multline}
It is clear that expression similar to (\ref{strun+1}) turns here into the following
\begin{equation}\label{strn+1}
p(t_{n},\dots,t_0 \mid s)=
\prod\limits_{i=0}^{n}p^0(t_i),\quad  s\in D_{n+1}.
 \end{equation}
Now, due to (\ref{strn}), (\ref{strn+1}) we have
\begin{multline}\label{pn}
p(t_{n},\dots,t_0)=\int\limits_0^\Delta p(t_{n},\dots,t_0 \mid s)f(s)ds=
\\=
\sum\limits_{k=0}^n
\prod\limits_{\vbox{\footnotesize\hbox{$i=0$}\hbox{$i\ne k$}}}^{n}p^0(t_i)
\int\limits_{D_k}
p\left(t_k \mid s-\sum\limits_{i=0}^{k-1} t_i\right)
g(s) ds+
\\+
\prod\limits_{i=0}^{n}p^0(t_i)\int\limits_{D_{n+1}}f(s) ds.
\end{multline}

\subsection{$t_0$-dependence cannot be eliminated in $p(t_{n+1} \mid t_n,\dots,t_0)$}

Now, with representations (\ref{pn+1}) for $p(t_{n+1},\dots,t_0)$ and
(\ref{pn}) for $p(t_{n},\dots,t_0)$ we can pose a question about the form
of $p(t_{n+1}\mid t_n,\dots, t_0)$. The latter can be found as defined in (\ref{defcond}).
First of all notice that due to (\ref{pn}) and Cond4,
$p(t_{n},\dots,t_0)$ is strictly positive for positive ISIs. 
This allows us to use it as denominator in the definition (\ref{defcond}).
Second, it can be further concluded from (\ref{pn}) and Cond4, that $p(t_{n},\dots,t_0)$ is
bounded, and therefore does not include any singularity of $\delta$-function type.
The latter means that any singularity contained in the $p(t_{n+1},\dots,t_0)$
appears as well in the $p(t_{n+1}\mid t_n,\dots, t_0)$. It follows from the above
that the conditional pdf $p(t_{n+1}\mid t_n,\dots, t_0)$ can be represented in the
following form:
\begin{equation}\label{firepr}
p(t_{n+1}\mid t_n,\dots, t_0) = p^w(t_{n+1}\mid t_n,\dots, t_0)+
Q(t_{n+1},\dots,t_0)\delta\left(\Delta-\sum\limits_{i=0}^{n+1}t_i\right),
\end{equation}
where $p^w(t_{n+1}\mid t_n,\dots, t_0)$ does not contain any $\delta$-function
depending on $t_0$, and $Q(t_{n+1},\dots,t_0)$ is strictly positive bounded function:
$$
Q(t_{n+1},\dots,t_0)=
\frac{a\prod\limits_{i=0}^{n+1}p^0(t_i)}{\lambda p(t_{n},\dots,t_0)}.
$$
The representation (\ref{firepr}) thus proves unequivocally that for any $n$,
conditional pdf $p(t_{n+1}\mid t_n,\dots, t_0)$ does depend on $t_0$ 
(the second term in (\ref{firepr})) 
and this dependence cannot be eliminated.

\section{Conclusions and Discussion}

We have proven here that any neuronal model, which satisfies Cond0-Cond4, above,
and is equipped with a delayed feedback, will display essentially non-Markov activity
expressed in terms of output ISIs, when stimulated with Poisson stream. 
This has a consequence for admissible approaches
while modeling activity of neuronal networks with stochastic behavior. Indeed, in
a reverberating network, a delayed feedback mediated by other neurons is always present.
Our result suggests that in this case, activity of individual neurons in the network
should be essentially non-Markov. Another situation in networks with instantaneous 
interneuronal communication. In the case of no delay communications, the neuronal
activity can well be Markov, or even Poisson, see example in \cite{Izhikevich2003}.

We used here a single neuron with delayed feedback as the simplest case of reverberating
"network". At the same time, neurons which send to themselves their output impulses
are known in real nervous systems, \cite{Nicoll1982,Bekkers1998}. Therefore, our conclusions
about essentially non-Markov behavior should be valid for those neurons even without taking
into account their involvement in a wider network activity.

The set of conditions Cond0-Cond4 while being rather natural and wide enough, leaves out
of our consideration many neuronal models known in neuroscience. E.g., Cond2 excludes
models with spike latency. Cond3 excludes models with internal memory extending beyond
a single ISI duration. 
Thus, we do not consider here 
partial afterspike resetting \cite{Rospars1993a,Lnsk1999}, threshold fatigue \cite{Chacron2003},
another types of adaptation, like multi-timescale adaptive threshold \cite{Kobayashi}.
Any kind of adaptation in individual neuron is by itself able to bring about a kind
of memory in the neuronal output stream. Therefore, considering neurons without adaptation
we demonstrate here, that delayed feedback without additional memory-like mechanisms 
known for neurons makes neuronal output essentially non-Markov.

Another limitation is Cond1 --- we use a Poisson process as a stimulus. It seems that
the proof given here can be extended to a wide class of renewal processes taken as stimuli.
This will be checked in further work.

\end{document}